# Coherent microwave generation by spintronic feedback oscillator

Dinesh Kumar[1], K. Konishi[2], Nikhil Kumar[3], S. Miwa[2], A. Fukushima[4], K. Yakushiji[4],
S. Yuasa[4], H. Kubota[4], C. V. Tomy[1], A. Prabhakar[3], Y. Suzuki[2], A. Tulapurkar[5]

[1] Department of Physics, Indian Institute of Technology Bombay, Powai, Mumbai – 400 076, India
[2] Graduate School of Engineering Science, Osaka University, Toyonaka, Osaka -560-8531, Japan
[3] Department of Electrical Engineering, Indian Institute of Technology Madras, Chennai –600036, India
[4] National Institute of Advanced Industrial Science and Technology (AIST), Spintronics Research Center, Ibaraki -305-8568, Japan
[5] Department of Electrical Engineering, Indian Institute of Technology Bombay, Powai, Mumbai – 400 076, India
*Correspondence to A.T. (ashwin@ee.iitb.ac.in)

The transfer of spin angular momentum to a nanomagnet from a spin polarized current provides an efficient means of controlling the magnetization direction in nanomagnets. A unique consequence of this spin torque is that the spontaneous oscillations of the magnetization can be induced by applying a combination of a dc bias current and a magnetic field. Here we experimentally demonstrate a different effect, which can drive a nanomagnet into spontaneous oscillations without any need of spin torque. For the demonstration of this effect, we use a nano-pillar of magnetic tunnel junction (MTJ) powered by a dc current and connected to a coplanar waveguide (CPW) lying above the free layer of the MTJ. Any fluctuation of the free layer magnetization is converted into oscillating voltage via the tunneling magneto-resistance effect and is fed back into the MTJ by the CPW through inductive coupling. As a result of this feedback, the magnetization of the free layer can be driven into a continual precession. The combination of MTJ and CPW behaves similar to a laser system and outputs a stable rf power with quality factor exceeding 10,000.

A spin polarized current, passing through a magnetic tunnel junction (MTJ) device exerts a spin transfer torque (STT) on the MTJ's free layer. The STT can, in turn, drive the magnetization of the free layer into continuous precession. Spin torque nano oscillators (STNOs) based on the STT[1] effect have attracted a considerable attention from the research community due to potential applications in wireless communication systems, ultra-sensitive magnetic field sensors and various other radio frequency (RF) devices[2-8]. Commendable work has been done on research and development of the STNO, especially to improve the output power and the quality factor of oscillations ($Q = f/\Delta f$). Georges et al. reported a large output power when a number of STNOs were electrically synchronized[9]. A mutual phase-locking of the STNOs was demonstrated to control an array of coupled STNOs, which resulted in an increased output power[10]. The output power can be also improved if one can reduce the critical current required to sustain oscillations[11]. The output power, from a single STNO, of up to 3.6 µW has thus far been reported[5, 12-16]. A maximum value of $Q = 7300$ ($f = 32.85$ GHz and $\Delta f = 4.5$ MHz) was reported in the case of a nanocontact based STNO[17]. The narrowest linewidth ($\Delta f$) of 280 kHz resulting in a $Q$ factor of 4000 has been reported for a spintronic vortex oscillator[18]. In the case of a MgO-based MTJ oscillator with in-plane magnetization, the maximum $Q$ value reported is 1000 with $f = 10$ GHz and $\Delta f = 10$ MHz[19]. The low $Q$ value in this case could be a result of incoherent oscillations across different parts of the free layer. In a different approach, S. Tamaru et al. have demonstrated extremely narrow line widths by developing a phase-lock loop (PLL) circuit specially designed for an STNO[20]. In this paper we demonstrate a dc current powered MTJ based RF oscillator, with a large quality factor, using an entirely different scheme[21]. Our scheme employs a co-planar waveguide (CPW) above the free layer of the MTJ. The voltage signal resulting from magnetization fluctuations in the free layer is amplified and fed to the CPW. The RF magnetic field generated around the CPW couples back to the free layer and enhances the resonant oscillations of the free layer. We observed highly coherent oscillations exhibiting linewidths as narrow as 200 kHz at ~ 2.5 GHz, resulting in a very large $Q$ factor of 12800, at room temperature. A spectrum of the power density also reveals interesting side bands accompanying the main peak, similar to the spectrum of a multimode laser.

We fabricated an MTJ stack on thermally grown $SiO_2$ (500 nm) with the following structure: bottom contact (50) / Ta(3) / Ru(5) / IrMn(7) / CoFe(3) / Ru(0.8) / CoFeB(3) / CoFe(0.4) / MgO(0.9) / CoFeB(3) /Ta(5) / Ru(5)/ top contact (45) (Fig 1(a)) (numbers in bracket denotes the thickness in nm) (see methods). The multilayer stack was patterned into elliptical nanopillars of size, $300 \times 500$ nm$^2$ using electron beam lithography and argon-ion milling methods. The easy axis of the free layer is taken to be along the x-axis; in-plane hard axis is along the y-axis and the out-of-plane hard axis is along the z-axis. The pinned layer magnetization is along the x-axis. All the layers are magnetized in-plane. A CPW was fabricated on top of the MTJ nano-pillar as shown in Fig. 1(a) and is electrically insulated from the MTJ by a 100 nm thick $SiO_2$ layer. The CPW is oriented in such a way that the current passing through it creates a magnetic field along the x-axis.

The resistance of the device, measured as a function of in-plane magnetic field applied at an angle of 45$^o$ with respect to the x-axis, is shown in Fig. S4 of the supplementary information. The device shows a high tunneling magneto-resistance (TMR) ratio of 50%. Figure 1(a) presents a schematic layout of the experimental set up for the measurements of the RF oscillations induced by the feedback effect. The external magnetic field is applied along the y-axis so as to obtain a non-collinear alignment of magnetic moments in the free and the fixed layers. A DC bias current was passed through the MTJ using a bias-T network. The RF port of the bias-T (feedback voltage) is connected to the CPW through a power splitter and an amplifier to amplify the feedback signal. (A phase shifter can also be inserted in the feedback path.) The CPW lies right above the free layer, and is electrically insulated from the MTJ by a thick (100 nm) $SiO_2$ layer. One part of the RF voltage, generated across the MTJ due to thermal fluctuations of the magnetization in the free layer, is measured by a spectrum analyzer, and the other part of it is amplified and fed into the CPW. The RF current passing through the CPW, creates an RF magnetic field along the x-axis. This RF magnetic field, which acts as the feedback, can amplify or suppress the fluctuations of the free layer magnetization which depends on the phase difference between the feedback signal and the magnetization oscillations of the free layer.

The measurements were performed on four samples – samples A, B, C and D, which were grown simultaneously under identical growth conditions. During the experiments the tunnel junction of MTJ broke down sometimes, possibly due to excessive dc bias current, electrostatic discharge and/or human error. Therefore, we had to use different samples for measurements. We obtained similar qualitative results from all four samples. We first measured the magnetic noise of the device (sample A) without connecting the feedback line by passing a dc current of -2 mA. The frequency of the peak in the noise spectra for various magnetic fields applied along the y-axis is shown in Fig. 1(b). The inset shows the noise spectra for H = 70 Oe and $I_{dc}$= -2 mA. According to Kittel's formula, the frequency depends on the magnetic field H, applied along y-axis as:

$$f = (\gamma/2\pi)\{H_c(H_c + H_d)[1-(H/H_c)^2]\}^{1/2} \quad H \leq H_c \qquad (1)$$
$$= (\gamma/2\pi)\{(H-H_c)(H-H_c+H_d)\}^{1/2} \quad H \geq H_c \qquad (2)$$

where $\gamma$ is the gyromagnetic ratio, $H_c$ and $H_d$ denote in-plane and out-of-plane anisotropy fields[8]. The frequency according to the above formula which is derived for single domain magnet, goes to zero at $H = H_c$. This feature shows up as a dip in the frequency in Fig 1(b). Micromagnetic simulations also reveal that frequency shows a dip as shown in Fig S1.

We then measured the power spectra of the device (sample A) by connecting the feedback line to the CPW. The gain of the amplifier, connected in the feedback circuit, was set to +24 dB. The power spectra for different dc bias currents ranging from -1.7 mA to -2.7 mA, in step of 0.1 mA, and H = 58 Oe are shown in Fig. 2(a). As we increase the current, the RF feedback magnetic field is increased (see equation 4 in supplementary information), which enhances the amplitude of oscillations of the free layer. For a bias current of -2.7 mA, we observed a large peak with a narrow line width of 200 KHz at ~ 2.5 GHz. This corresponds to a significant increase in the quality factor of the oscillations ($Q \sim 12800$). The variation of the frequency and the line width as a function of dc bias current is plotted in Fig. 2(b). Figure 2(c) shows the total power output as a function of bias current. The power output increases steeply with increasing current. The power spectra obtained for the same values of currents and magnetic field, but without feedback are shown in Fig S6. Thus though it is possible that STT can affect the experimental results, the comparison of figures 2(a) and S6 shows that the feedback effect is responsible for high power output and quality factor.

Figure 3(a) shows the power spectra of data shown in Fig. 2(a), for dc current values varying between -2.2 mA to -2.7 mA, on a log scale. We can clearly see the side peaks around the main peak. The frequency difference between the side peaks is about 120 MHz. Our system is similar to a multi-mode laser system, which also falls under the category of a delay-line oscillator. The delay in the case of laser system is due to the optical cavity[22], whereas in the present experiment, the feedback line provides the delay. The difference between the side peaks is determined by the round trip delay of the system. The observed frequency difference of 120 MHz would correspond to a delay of about 8.3 ns.

Figures 3(b), 3(c) and 3(d) show the power spectra obtained on sample B with H = 92 Oe and $I_{dc}$ = +1 mA, for various values of the gain of amplifier. An increase in feedback gain is accompanied by an increase in the amplitude of magnetic oscillations. Figure 3(b) shows the power spectra for the case without feedback (red curve) and with an amplifier gain of +10 dB (green curve). On connection of the feedback circuit we begin to observe side peaks adjacent to the main resonant peak in the spectrum. On further increase of the amplifier gain to +20 dB, the side peaks are more evident as shown in Fig. 3(c). For amplifier gain = +33 dB (as shown in Fig. 3(d)), the intensity of the fundamental peak becomes very high and consequently the side peaks are not visible. We also observe a reduction in the linewidth with an increase in the amplifier gain. In the absence of feedback, from the data shown in figure 3(b), the linewidth obtained from Lorentzian fitting is 387 MHz, with oscillations in the free layer magnetization reaching a maximum of ~ 2 pW/MHz. For identical biasing conditions, but now with an amplifier gain of +33 dB (as shown in Fig. 3(d)), the linewidth obtained was 10 MHz, with oscillations in the free layer magnetization reaching a maximum of ~ 2.5 nW/MHz. Thus, the RF feedback has increased the amplitude of oscillations by three orders of magnitude. (The corresponding increase in the total power is by about a factor of 32).

The above results show that the feedback effect is the dominant factor in our experiments as compared to the STT effect. The amplified signal is not fed back into the MTJ directly, but rather coupled to the MTJ through the inductive coupling of feedback line. Thus any STT effect present in MTJ is not amplified and cannot give rise to the results shown in fig 3(b)-3(d). The critical voltage required for STT excitation is estimated to be 1.18 V which is much larger than maximum voltage of 0.27 V used in these experiments (see supplementary information). Further it should be noted that if we use an amplifier to simply amplify the noise signal (*e.g.* spectrum without feedback shown in fig 3(b)), we will get a larger output power, however, the line width would be the same. The amplifier in the present experiment substantially reduces the line width due to the feedback effect.

We also carried out experiments by inserting an additional cable of about 1.5 m length in the feedback path. The power spectral density (PSD) measured (on Sample C) with a current of -2 mA (amplifier gain = 27 dB, H = 70 Oe) is shown in Fig. S5 of the supplementary information. The peak separation decreased to 62.2 MHz which corresponds to a round trip delay of 16.1 ns.

The combination of an MTJ and a CPW provides a gain to the microwaves just like light is amplified in an optical gain medium. There are, however, differences in the physical mechanism of amplification. In case of the laser, light is amplified by stimulated emission process, as shown schematically in Fig. 4(a). An atom in excited state makes a transition to a lower energy state due to incident photon, and emits another photon. The transition is typically driven by the electric field of the incident electromagnetic wave. A similar schematic diagram can be drawn in the case of feedback amplification as shown in Fig. 4(b). If a microwave is incident on the CPW, the magnetic field associated with the incident wave can excite oscillations of the free layer magnetization. The dc current flowing through the MTJ, converts these oscillations into microwaves via the TMR effect. For large TMR ratios (or large dc currents), we can amplify the incident microwave, which is shown schematically as emission of two microwave photons from the device in response to one incident photon. The microwaves emitted by the device have a fixed phase relative to the incident microwaves. When we connect the CPW and the MTJ, as shown in Fig. 1(a), any fluctuation of the free layer magnetization results in a fluctuating current signal in the CPW. This acts as incident microwave radiation which gets amplified and finally results in the emission of a coherent microwave radiation. In this experiment we had to use an external amplifier as the gain of the system was low due to the small TMR ratio and large width of the feedback line.

We have carried out micromagnetic simulations of our device to gain further insight into the feedback process. We simulated an elliptical sample of 500 nm x 300 nm x 3 nm dimensions using MuMax3 (details in supplementary information). Figure 5(a) shows the power spectral density of $m_x$ for various currents assuming an amplifier gain of +21 dB, H = 60 Oe and a feedback delay of 10 ns. The curve with $I = 0$, corresponds to the case of no feedback, and shows the thermal fluctuations of the free layer. One can clearly see that as we increase the current, the peak in the PSD grows in amplitude and becomes sharper, similar to the trend observed in the experiments. In Fig. 5(b), we have plotted the cross-correlation of $m_x$ (at zero lag) between the centre of the sample and various points along the long axis of the sample. For low currents we see that the cross-correlation drops quickly to zero as we move towards the sample edges. As we increase the current, the cross correlation improves. At high currents, where we see the large amplitude peaks in the PSD, the cross correlation is close to 1 across the sample. This implies that at high current the magnetization oscillations of various parts of the sample are phase-locked. The feedback signal depends on the resistance of the sample, *i.e.*, the average value of $m_x$ of the entire free layer. As the feedback signal contains the information about the average value of $m_x$, it induces oscillations with the same phase across the sample. This, in turn, implies that the entire sample can oscillate coherently like a single domain particle. The micromagnetic simulations further reveal the presence of side bands as observed in the experiment (supplementary Fig. S2). The PSD for delays of 10 and 20 ns are shown in Fig. S2. The distance between the side bands for delay of 10 ns and 20 ns is about 0.1 GHz and 0.05 GHz, respectively, *i.e.*, equal to the inverse of delay time.

We also analyzed the present oscillator in terms of a universal oscillator model[10] by incorporating the feedback effect. We found that the feedback can significantly reduce the line width of the oscillator as expressed by the equation below:

$$FWHM = \frac{\alpha kT}{4\pi p_0 (1-p_0) S} \left( 1 + \left( \frac{\Gamma_p}{\Gamma_p + p_0 \omega_f} \right)^2 \nu^2 (1-p_0)^2 \right) \frac{1}{(1 + \omega_f \Delta t (1-p_0))^2} \quad (3)$$

In the above equation ν denotes dimensionless frequency shift *w.r.t.* oscillation amplitude, $p_0$, α is the Gilbert damping constant and T is temperature. $\Gamma_P$ is the dynamic damping factor. The details are given in supplementary note 3.

We further carried out the PSD measurements on sample D at higher amplifier gains to get more power output. The results of such a measurement, in a magnetic field of 70 Oe and with an amplifier gain of +29 dB, are shown in Fig. 6 for both polarities of the dc bias current. One can see that there is a difference between the power output for positive and negative currents. Changing the polarity amounts to changing the phase by 180°, which can give rise to this asymmetry (see supplementary note 6). Spin transfer torque could also contribute to this asymmetry. We have seen this asymmetric behavior in all the samples tested. We could obtain a large power output of more than one microwatt for positive currents. From Fig. 6, we can also see that there is a threshold current of about +1.5 mA, above which the power output rises steeply. Similar results have been obtained from the micromagnetic simulations as shown in Fig. S3 of the supplementary information. Above this threshold current (see Fig. S3), we can also see a large enhancement in the cross-correlation function as shown in Fig. 5(b). This implies that above this threshold value of the current, different parts of the sample oscillate in-phase and we get a large output power.

We have used an amplifier with a gain of +29 dB in the feedback path to get a large power output. As the signal from the MTJ is split into 2 equal parts (Fig. 1(a)), the signal is effectively amplified by +26 dB, which corresponds to a voltage gain of about 20. The RF feedback magnetic field can be increased by decreasing the width of feedback line and increasing the TMR ratio of the device (see equation 4 in supplementary information). The present experiment has a feedback line width of 1 µm and a TMR of 50%. Decreasing the width to 100 nm and increasing the TMR ratio to 200 % would provide a gain of 40, making the use of an external amplifier completely unnecessary. We have also restricted the measurements to an applied dc bias of about 0.27 V. A further increase in the feedback amplification can be obtained by increasing the dc bias current.

In conclusion, we have demonstrated coherent microwave emission from a magnetic tunnel junction by using magnetic field feedback. A large quality factor, exceeding 10000, was obtained experimentally. These nanoscale oscillators would find several applications such as in wireless communication systems. One of the features of the present oscillator configuration is that the spin transfer torque is not required for its operation. Devices with interplay of both spin torque and feedback effect, could lead to novel physical effects and better oscillators.

**Methods:**

Multilayers with the stacking structure of bottom contact (50) / Ta(3) / Ru(5) / IrMn(7) / CoFe(3) / Ru(0.8) / CoFeB(3) / CoFe(0.4) / MgO(0.9) / CoFeB(3) / Ta(5) / Ru(5)/ top contact (45) (thickness in nanometers) were fabricated. The MTJ film was deposited using magnetron sputtering using a Canon ANELVA C7100. The sample was post-annealed at 300$^o$ C for 2 hours in an in-plane field of 6 kOe. CoFe(3)/ Ru(0.8)/ CoFeB(3) is the synthetic antiferromagnetic (SAF) polarizing layer. CoFe and CoFeB are coupled antiferromagnetically through Ru. Top CoFeB layer acts as a free layer. Ta(5)/Ru(5) is the capping layer for CoFeB free layer which is made quite thick to protect the free layer from any kind of damages during the microfabrication processes. The microwave emission spectra were measured with a spectrum analyzer. In order to obtain the correct RF emitted output power from the MTJ, we have subtracted the background data from the raw output power data. Each spectrum is obtained by averaging 100 spectral scans. The power output has been corrected for the inclusion of the power splitter in the system and the impedance mismatch. All measurements were carried out at room temperature. The micromagnetic simulations have been performed with the MuMax3 program [23].

**Figure captions**

Figure 1: (a) Schematic diagram of the feedback oscillator. The top layer of the MTJ pillar shows the free layer, middle layer shows the tunneling barrier, and the bottom layer shows the pinned layer. A coplanar waveguide (CPW) rests on top of the free layer and is electrically insulated from the MTJ. A DC current is passed through MTJ via Bias-T. The oscillating voltage produced across the MTJ, due to the oscillations of free layer magnetization, is split into two paths using a power splitter. One part is amplified, using an amplifier in the feedback circuit, and fed into the CPW. The second part of oscillating voltage is observed on the spectrum analyzer. The oscillating current in the CPW creates an ac magnetic field on the free layer, which acts as the feedback. The phase between the free layer magnetization oscillation and the ac magnetic field can be adjusted by the phase shifter. (b) Frequency of the peak in the noise spectrum as a function of magnetic field applied along the y-direction, for $I_{dc}$ = -2 mA. The inset shows the noise spectrum obtained for H = 70 Oe and $I_{dc}$ = -2 mA. The noise spectra were measured by disconnecting the feedback waveguide. The data is taken for sample A.

Figure 2: (a) Power spectral density (PSD) as a function of dc current ranging from -2.2 mA to -2.7 mA, with an applied magnetic field of 58 Oe along the *y*-axis. The amplifier gain was set to +24 dB. Inset shows the power spectral density for low dc current values, ranging between -1.7 mA to -2.1 mA. As the dc bias current increases, the peaks grow in amplitude and become sharper. The PSD for different currents shown in the inset are multiplied by various factors for clear visibility. (b) Variation of frequency and line width as a function of bias current. The narrow line width of 200 kHz obtained at -2.7 mA corresponds to a quality factor of ~12800. (c) The total power output as a function of dc bias current. The data is taken for Sample A.

Figure 3: (a) Power spectra of the plots shown in Fig. 2(a) in log scale. The side peaks can be clearly seen. (b) Power spectral density for the case of no feedback (red curve) and for case when gain of the amplifier is +10 dB (green curve). The green curve has been shifted vertically upwards for clarity. It shows some side peaks along with the fundamental peak. (c) Power spectral density when gain of the amplifier is +20 dB. The side peaks are more evident in this case. (d) Power spectral density for amplifier gain of +33 dB. In this case the intensity of fundamental peak is enhanced greatly and consequently the side peaks are not visible. The linewidth of the peak decreases with increasing gain. The data for graphs 3(b), 3(c) and 3(d) is taken using sample B with H=92 Oe and $I_{dc}$=1 mA

Figure 4: Comparison of the amplification processes in a laser and feedback oscillator. (a) Schematic diagram of amplification of photons by stimulated emission in a laser (b) Schematic diagram of the amplification of microwave photons by a combination of MTJ powered by a dc bias current and a co-planar wave guide (CPW). A microwave photon incident on the CPW, excites magnons in the free layer. The magnons in the free layer, generate microwaves due to the TMR effect and a dc bias current. If the TMR effect is large (or if dc bias current is large), the incident microwaves can be amplified.

Figure 5: Simulation results: (a) The spectral density of $m_x$ for various dc current values. A magnetic field of 60 Oe was applied along $y$ axis and amplifier with gain of +21 dB was assumed in the feedback line. As the dc bias current is increased the peaks in the spectral density grow in amplitude and become sharper indicating improvement in the linewidth of the peak. (b) The magnetization of each cell on the major axis of the ellipse was recorded as a function of time. The cross-correlation (at 0 lag) between the $m_x$ at the center and $m_x$ along the axis is plotted for different values of dc bias current. For low values of bias currents, the cross-correlation as a function of distance decays to zero rapidly. For larger bias currents, where the amplitude of peak in spectral density is large, the cross-correlation remains large even near the sample edges. Thus the entire sample oscillates coherently for large currents.

Figure 6: Power output as a function of dc bias current for amplifier gain of + 29 dB and H = 70 Oe. The data is taken using sample D.

**Acknowledgment:**

We would like to acknowledge the support of Center of Excellence in Nanoelectronics (CEN) and IITB-NF facility in Department of Electrical Engineering, IIT Bombay. DK would like to acknowledge UGC, New Delhi for the financial support through senior research fellowship during the research work. We would also like to acknowledge Swapnil Bhuktare for help during the measurements. This work was partly supported by Grant-in-Aid for Scientific Research (B: 16H03850) of JSPS.

**Author Contributions:** MTJ film was deposited by KK with supervision from AF, KY, SY and HK. Device fabrication was done by KK with supervision from SM and YS. DK carried out the measurement, analysed the data and wrote the manuscript with help from AT and CVT. AT designed the experiment and conceived the project with inputs from YS. NK and AP were involved in the micromagnetic simulation and analysis. YS carried out analytical modelling of feedback oscillator. All authors contributed to this work and commented on this paper.

**Competing Financial Interests:** Authors declare that they have no Competing Financial Interests.

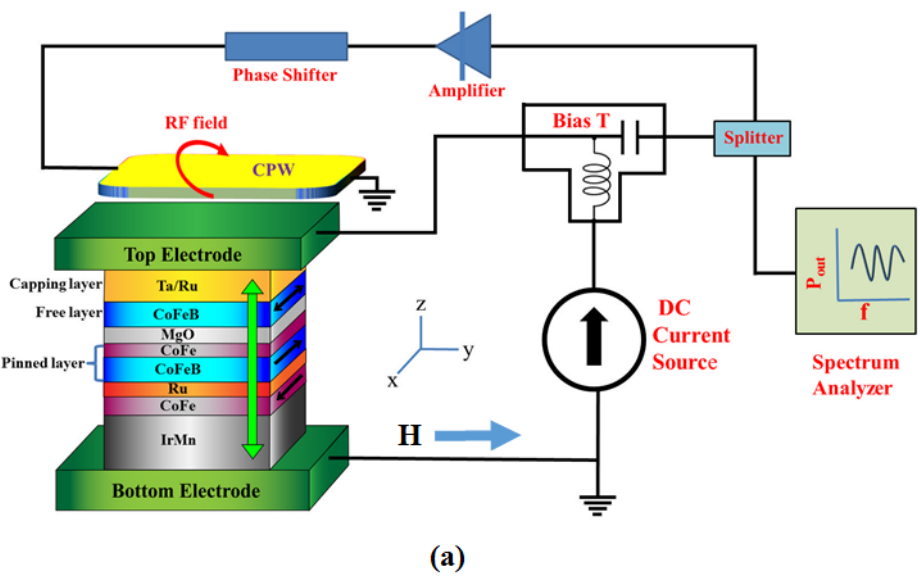 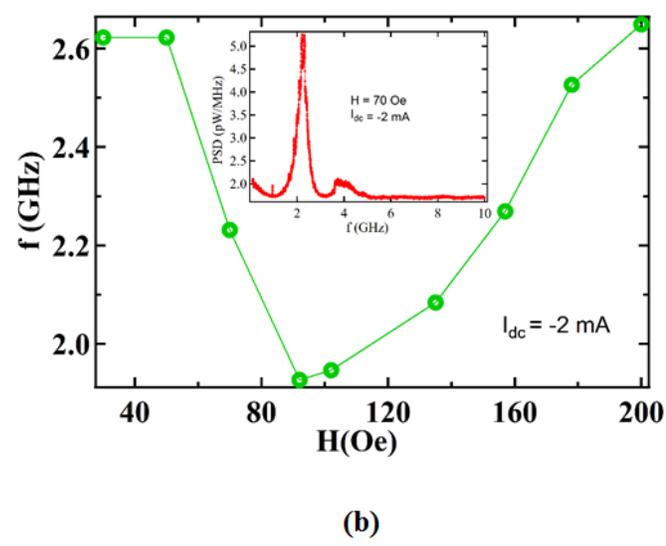

(a) (b)

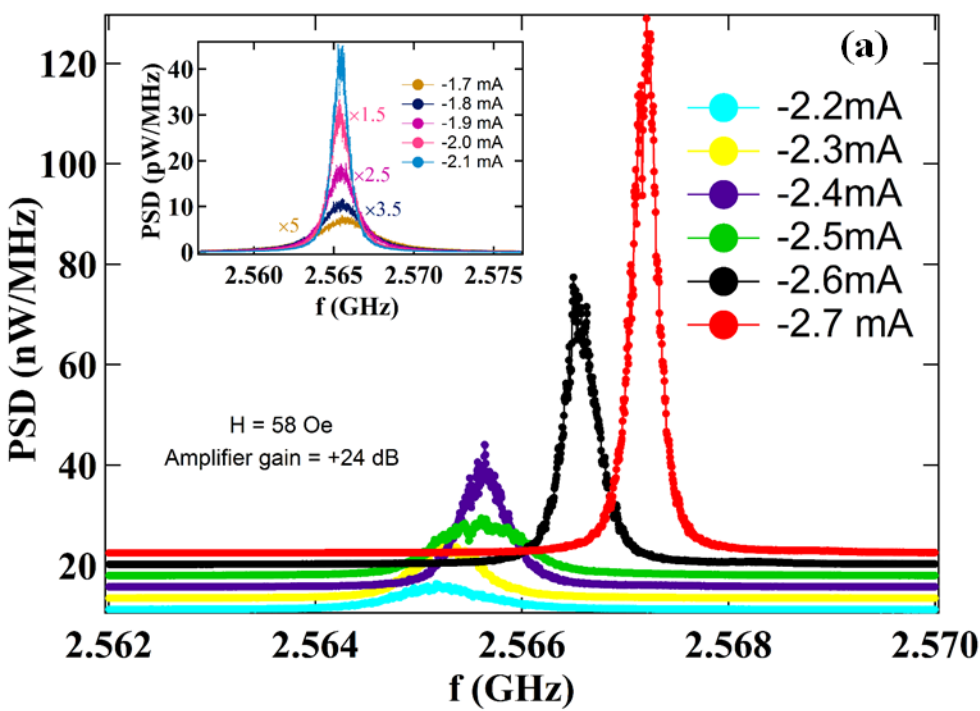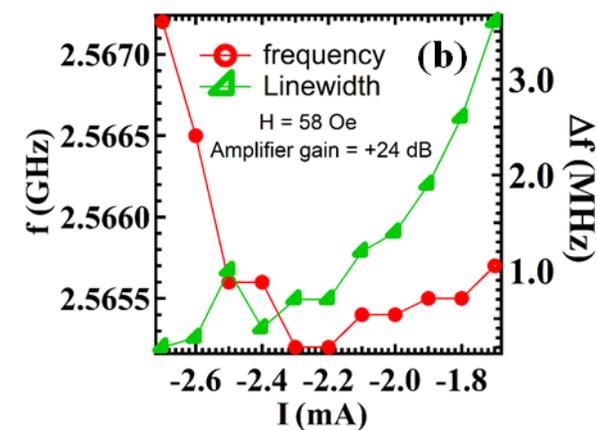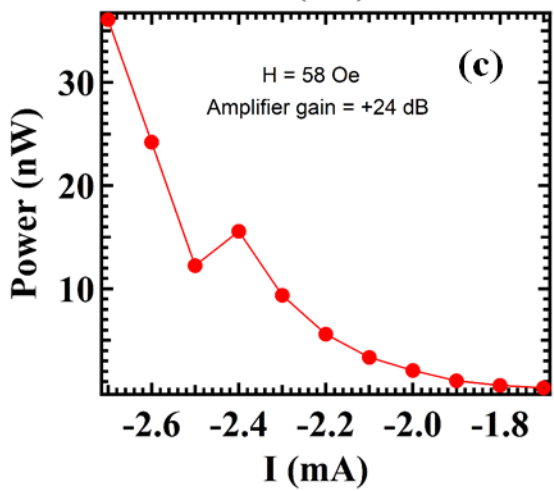

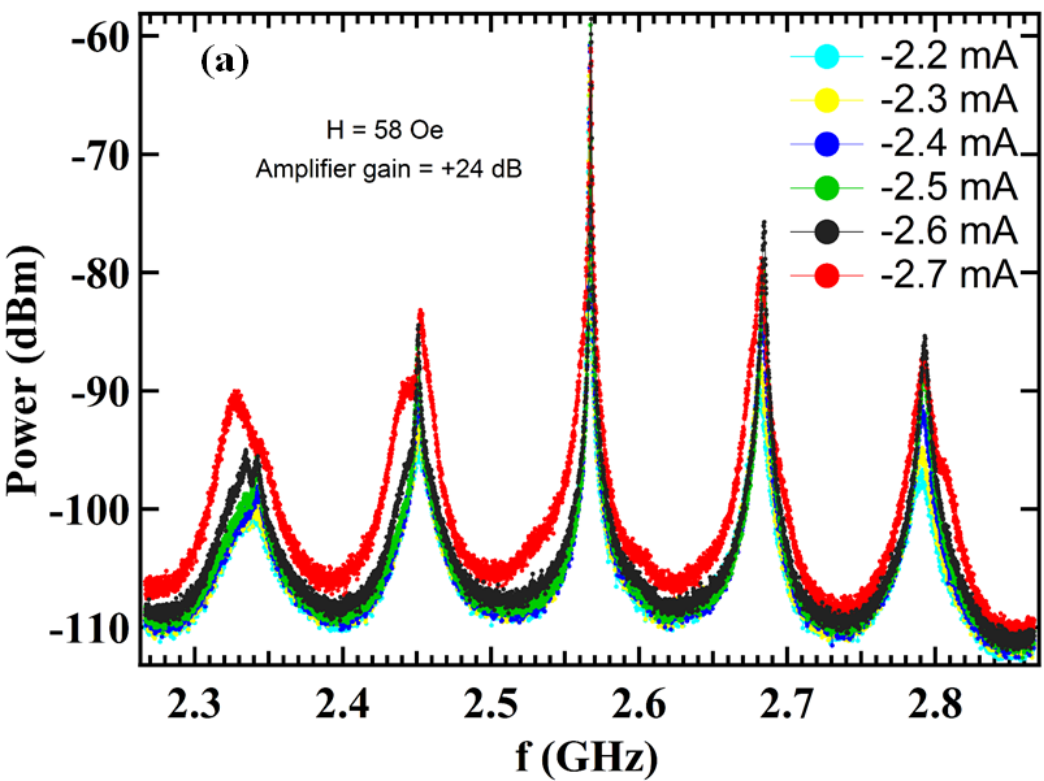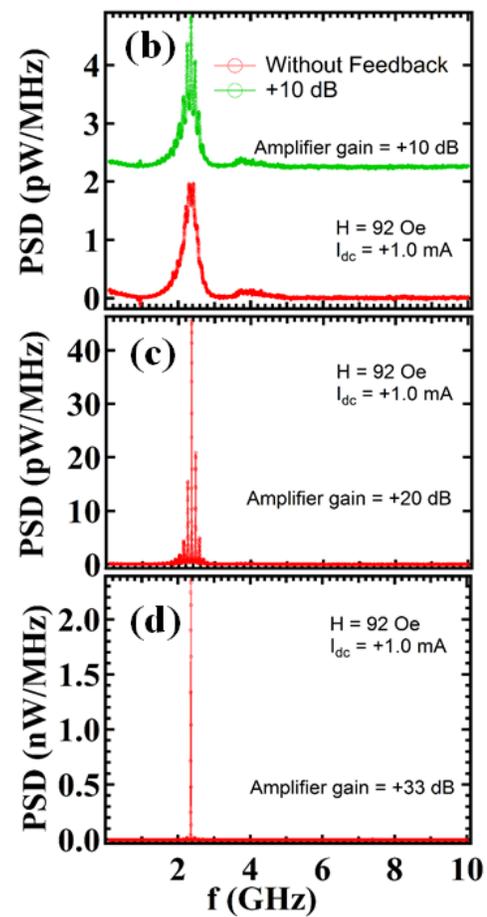

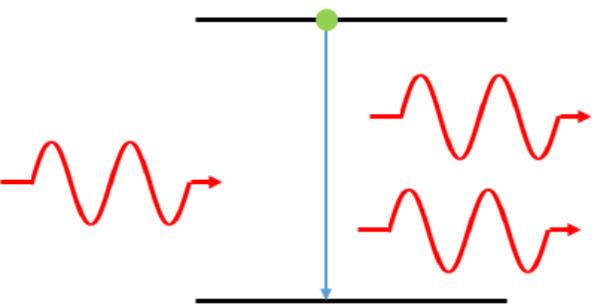 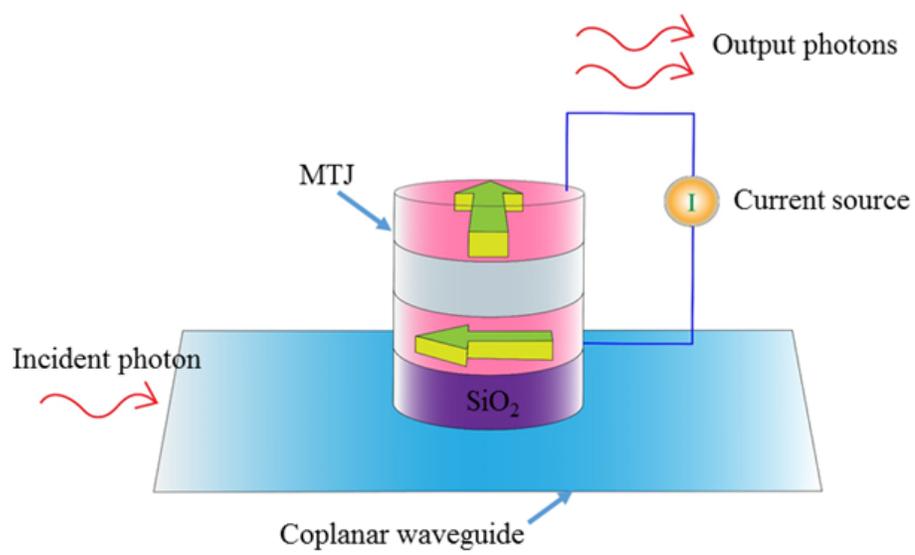

(a) (b)

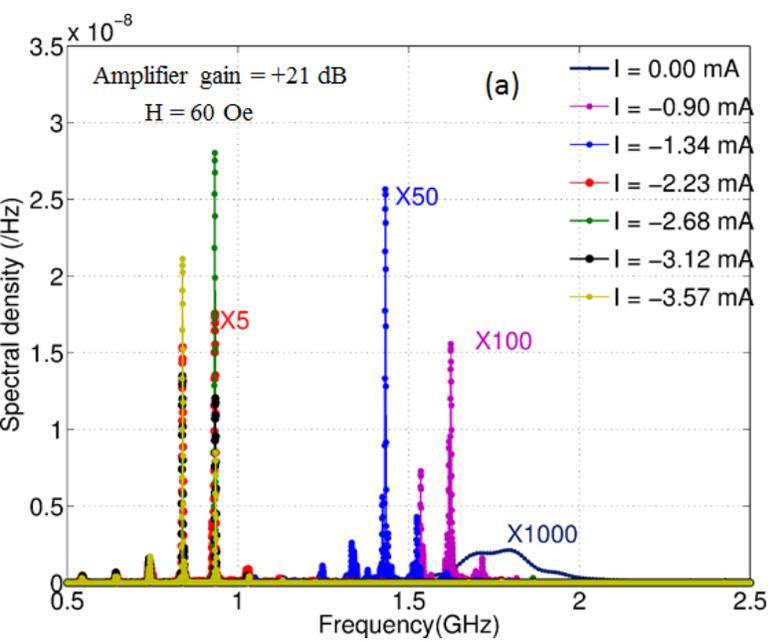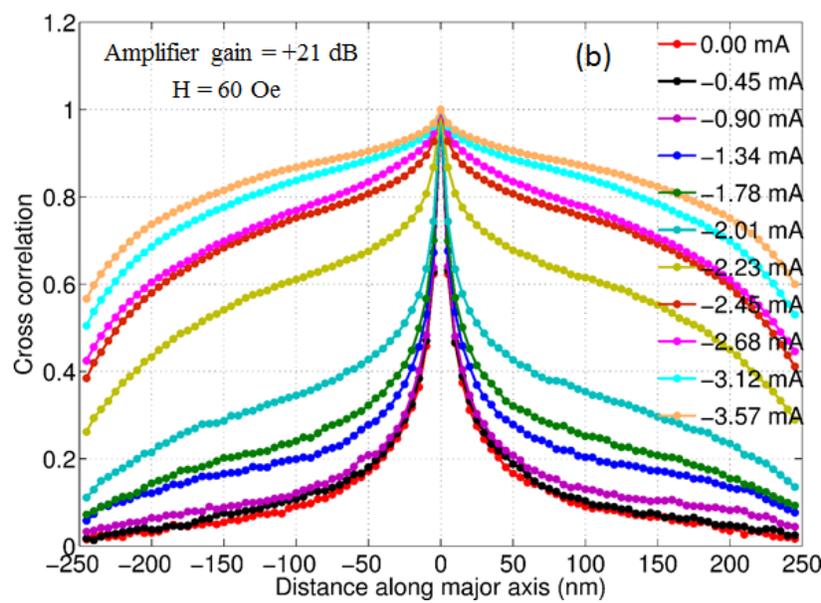

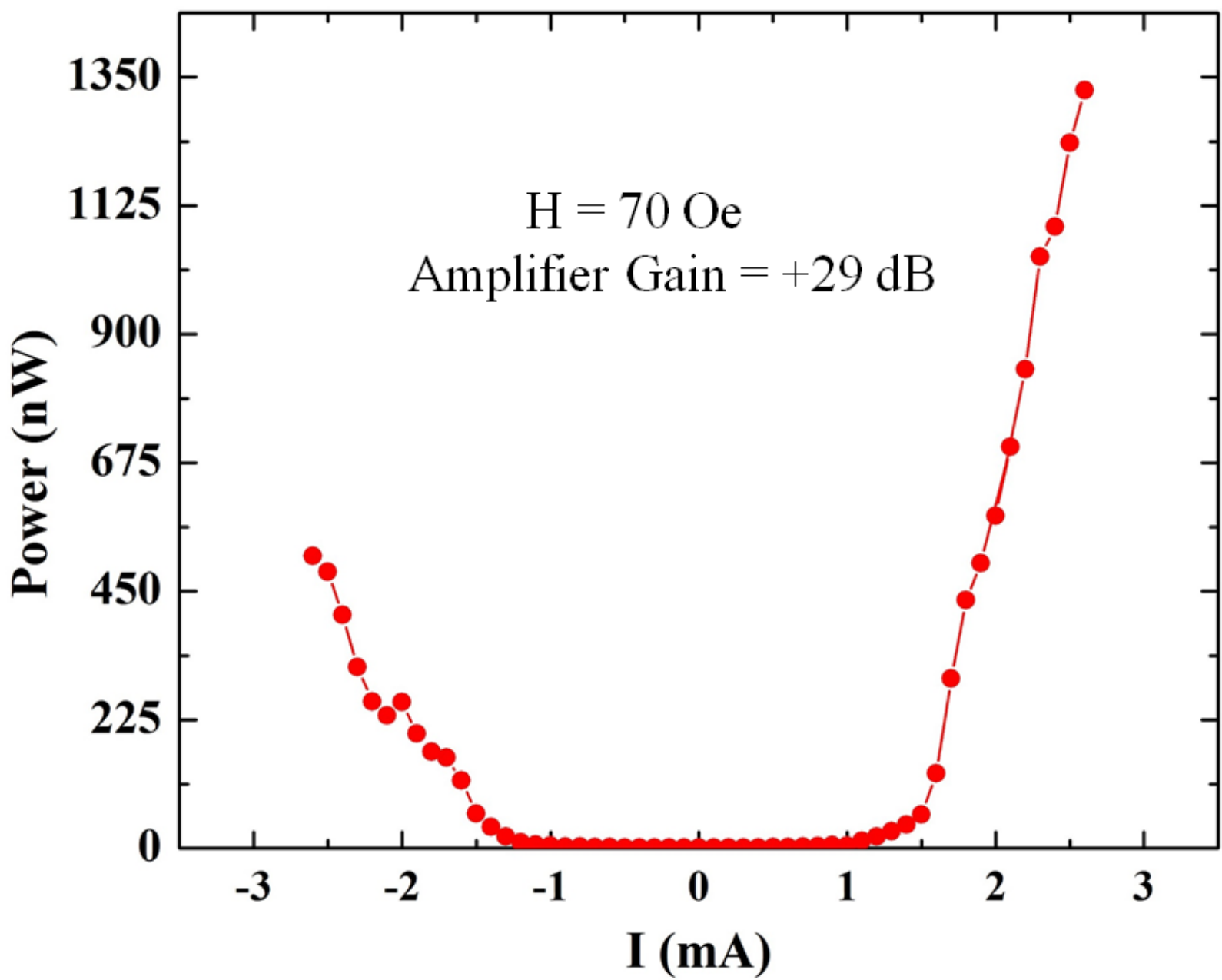